\documentclass{llncs}

\pdfoutput=1

\usepackage{latexsym}
\usepackage{pstricks,pst-node,epsfig}
\usepackage{eucal}

\usepackage{amssymb}

\newcommand{\be}{\begin{eqnarray}}
\newcommand{\ee}{\end{eqnarray}}
\newcommand{\bes}{\begin{eqnarray*}}
\newcommand{\ees}{\end{eqnarray*}}

\newcommand{\scr}{\EuScript}
\DeclareMathSymbol{\upset}{\mathopen}{symbols}{"22}
\DeclareMathSymbol{\downset}{\mathopen}{symbols}{"23}
\newcommand{\myA}{\scr A}
\newcommand{\myB}{\scr B}
\newcommand{\myF}{\scr F}

\newcommand{\binsum}[2]{{~#1 \choose \downset#2}}
\newcommand{\ds}[2]{#1 \boxtimes #2}

\newcommand{\nO}{O}
\newcommand{\no}{o}
\newcommand{\Ostar}{O^{*}}

\begin{document}

\pagestyle{headings}

\title{Counting Paths and Packings in Halves}

\author{
	Andreas Bj\"orklund\inst{1} 
	\and 
	Thore Husfeldt\inst{1,2} 
	\and 
	Petteri Kaski\inst{3}\thanks{
	This research was supported in part by the Academy of Finland, 
	Grants 117499 (P.K.) and 125637 (M.K.).}
	\and 
	Mikko Koivisto\inst{3}\inst{\star}
}

\institute{
	Lund University,
	Department of Computer Science,\\
	P.O.Box 118, SE-22100 Lund, Sweden\\
	\email{andreas.bjorklund@yahoo.se},
	\email{thore.husfeldt@cs.lu.se}
        \and
        IT University of Copenhagen,\\
        Rued Langgaards Vej 7, 2300, K\o{}benhavn S, Denmark
	\and
	Helsinki Institute for Information Technology HIIT,\\
	Department of Computer Science, University of Helsinki,\\
	P.O.Box 68, FI-00014 University of Helsinki, Finland\\
	\email{petteri.kaski@cs.helsinki.fi},  
	\email{mikko.koivisto@cs.helsinki.fi}
}

\maketitle

\begin{abstract}
It is shown that one can count $k$-edge paths in an $n$-vertex graph 
and $m$-set $k$-packings on an $n$-element universe, respectively,  
in time ${n \choose k/2}$ and ${n \choose mk/2}$, up to a factor 
polynomial in $n$, $k$, and $m$; in polynomial space, the bounds hold if 
multiplied by $3^{k/2}$ or $5^{mk/2}$, respectively. These are 
implications of a more general result: given two set families on an 
$n$-element universe, one can count the disjoint pairs of sets in 
the Cartesian product of the two families with $\nO(n \ell)$ basic 
operations, where $\ell$  is the number of members in the two families 
and their subsets.
\end{abstract}

\section{Introduction}

Some combinatorial structures can be viewed as two halves that meet in the
middle. For example, a $k$-edge path is a combination of two $k/2$-edge paths. 
{\em Bidirectional search} \cite{Danzig,Pohl} finds such structures by searching
the two halves simultaneously until the two search  frontiers meet. In
instantiations of this idea, it is crucial to efficiently {\em join} the two
frontiers to obtain a valid or optimal solution.  For instance, the
meet-in-the-middle algorithm for the Subset Sum problem,  by Horowitz and Sahni
\cite{HorowitzSahni},  implements the join operation via a clever pass through
two sorted lists of subset sums.  

In the present paper, we take the meet-in-the-middle approach to counting
problems, in particular, to counting paths and packings.  Here, the join
operation amounts to consideration of pairs of {\em disjoint} subsets of a
finite universe, each subset weighted by the number of structures that span the
subset. We begin in Sect.~2 by formalizing this as the {\em Disjoint Sum} 
problem and providing an algorithm for it based on inclusion--exclusion
techniques \cite{BHK,BHKK-stoc,BHKK-stacs,Karp,Kennes,KohnGottliebKohn}.  In
Sect.~3 we apply the method to count  paths of $k$ edges in a given $n$-vertex
graph in time  $\Ostar\big({n \choose k/2}\big)$; throughout the paper,
$\Ostar$  suppresses a factor polynomial in the mentioned parameters (here, $n$
and $k$). In Sect.~4 we give another application, to count $k$-packings in a
given family of $m$-element subsets of an $n$-element universe in time 
$\Ostar\big({n \choose mk/2}\big)$.  For both problems we also present  slightly
slower algorithms that require only polynomial space. 

We note that an earlier report on this work under a different title 
\cite{BHKK-corr} already introduces a somewhat more general technique and an
application to counting paths. The report 
has been cited in some recent papers
\cite{AlonGutner,VassilevskaWilliams}, which we, among other related previous
work, discuss below.

\subsection{Related Work and Discussion}

Deciding whether a given $n$-vertex graph contains a Hamiltonian path, that is,
a simple path of $n-1$ edges, is well known to be NP-hard. The fastest known
algorithms, due to Bellman \cite{Bellman,Bellman2} and, independently,  Held and
Karp \cite{HeldKarp}, are based on dynamic programming across the vertex subsets
and run in time $\Ostar(2^n)$.  Equally fast polynomial-space variants that
actually count all Hamiltonian paths via inclusion--exclusion were discovered
later by Kohn, Gottlieb, and Kohn  \cite{KohnGottliebKohn}, and independently,
Karp \cite{Karp}. Our algorithm (cf.\ Theorem~\ref{thm:path}, for $k=n-1$), too,
runs in time $\Ostar(2^n)$, if allowing exponential space. 

In this light, it is intriguing  that the parameterized problem of counting
paths of $k$ edges seems harder than the corresponding decision problem; this is
the present understanding  that has emerged from a series of works, starting
perhaps in Papadimitriou and Yannakakis's \cite{PapadimitriouYannakakis} 
conjecture that for  $k = O(\log n)$ the decision problem can be solved in
polynomial time.  The conjecture was proved by Alon, Yuster, and Zwick's
\cite{AlonYusterZwick} color-coding technique that  gave a randomized algorithm
with expected running time $\Ostar(5.44^k)$ and a derandomized variant with
running time $\Ostar(c^k)$ for a large constant $c$. With a more efficient
color-coding scheme, Chen, Lu, Sze, and Zhang \cite{Chen_et_al} improved the
latter bound to $\Ostar(12.8^k)$; see also Kneis, M\"{o}lle, Richter, and
Rossmanith \cite{Kneis_et_al}.  Using completely different techniques, Koutis
\cite{Koutis2008}, followed by Williams \cite{Williams},  developed a randomized
algorithm that runs in expected time $\Ostar(2^k)$.  Unfortunately, it is
unlikely that the randomization based techniques extend to counting. For
instance,  very recently Alon and Gutner \cite{AlonGutner} showed that
color-coding is doomed to fail as every ``balanced'' family of hash functions
from a $k$-set to an $n$-set is of size at least  $c(k) n^{\lfloor k/2 \rfloor}$
for some function $c$. Flum and Grohe \cite{FlumGrohe} proved another negative
result, namely that the counting problem is \#W[1]-hard with respect to the
parameter $k$.  From a positive side, a very recent result of Vassilevska and
Williams \cite{VassilevskaWilliams}  implies that $k$-edge paths can be counted
in time  $\Ostar\big(2^k (k/2)!{n \choose k/2}\big)$ in polynomial space;  our
polynomial-space algorithm (cf.\ Theorem~\ref{thm:polypath}) is faster still, by
a factor of $(4/3)^{k/2} (k/2)!$.

Concerning set packings the situation is analogous,  albeit the research has
been somewhat less extensive. Deciding whether a given family of $f$ subsets of
an $n$-element  universe contains a $k$-packing is known to be W[1]-hard
\cite{DowneyFellows}, and thus it is unlikely that the problem is fixed
parameter tractable, that is, solvable in time $c(k) f^d$ for some function $c$
and constant $d$.  If $f$ is fairly large, say exponential in $n$,  the fastest
known algorithms actually count the packings by  employing the
inclusion--exclusion machinery  \cite{BHK,BHKK-stoc} and run in time
$\Ostar(2^n)$.  This bound holds also for the presented algorithm (cf.\
Theorem~\ref{thm:pack}).

Again, it is interesting that there is a natural parameterization under which
counting $k$-packing seems harder than the corresponding decision problem. 
Indeed, Jia, Zhang, and Chen \cite{JiaZhangChen} showed that the decision
problem is fixed parameter tractable with respect to the total size $mk$ of the
packing, assuming each member is of size $m$. Koutis \cite{Koutis2005}, followed
by  Chen, Lu, Sze, and Zhang \cite{Chen_et_al}, gave faster algorithms  with
running time $\Ostar(c^{mk})$ for some constant $c$; we note that here the
running time also grows about linearly in the input size $f$, which can be as
large as ${n \choose m}$. For {\em counting} $m$-set $k$-packings, previous
techniques \cite{BHK,BHKK-stoc} alone only give a running time bound of 
$\Ostar\big({n \choose mk}\big)$ if $mk \leq n/2$ and $\Ostar(2^n)$ otherwise.
Besides the present work, we are aware of two recent improvements: 
For the special case of counting $t$-matchings, that is $2$-set
$t/2$-packings,\footnote{Whether counting $t$-matchings is fixed-parameter tractable
remains a major open question in parameterized complexity.} 
Vassilevska and Williams \cite{VassilevskaWilliams} give a
time bound of $\Ostar\big(2^{t+c(t)} {n \choose t/2}\big)$, where $c(t)$ is of
the order $\no(t)$; our polynomial-space algorithm (cf.\
Corollary~\ref{cor:m2}) turns out to be slightly faster, by a factor of about
$(4/3)^{t/2}$. For the general case, Koutis and Williams \cite{KoutisWilliams}
give a time bound of $\Ostar\big(n^{mk/2}\big)$; our bounds 
(Theorem~\ref{thm:pack}) appear to be superior, e.g., when $mk$ grows
linearly in $n$.

The presented meet-in-the-middle approach resembles the randomized
divide-and-conquer technique by Chen, Lu, Sze, and Zhang \cite{Chen_et_al} and
the similar divide-and-color method by Kneis, M\"{o}lle, Richter, and Rossmanith
\cite{Kneis_et_al}, designed for parameterized decision problems. These can, in
turn, be viewed as extensions of the recursive partitioning technique of Gurevich
and Shelah \cite{GurevichShelah} for the  Hamiltonian Path problem. That said,
our contribution is rather in the observation that, in the counting context, the
join operation can be done efficiently using the inclusion--exclusion
machinery. While our formalization of the problem as the Disjoint Sum problem is
new, the solution itself can, in essence, already be found in Kennes
\cite{Kennes},  even though in terms of possibility calculus and without
the idea of ``trimming,'' that is, restricting the computations to small
subsets. Kennes's results were rediscovered in a dual form and extended to
accommodate trimming  in the authors' recent works
\cite{BHK,BHKK-stoc,BHKK-stacs}.

\section{The Disjoint Sum Problem} 

Given two set families $\myA$ and $\myB$, and functions $\alpha$ and $\beta$
that associate each member of $\myA$ and $\myB$, respectively, an element from a
ring $R$, the {\em Disjoint Sum} problem is to find the sum of the products
$\alpha(A) \beta(B)$ over all {\em disjoint} pairs of subsets $(A, B)$ in the
Cartesian product  $\myA \times \myB$; denote the sum by $\ds{\alpha}{\beta}$.
In applications, the ring $R$ is typically the set of integers equipped with the
usual addition and multiplication operation.  Note that, had the condition of
disjointness removed, the problem could be easily solved using about
$|\myA|+|\myB|$ additions and one multiplication.  However, to respect the
disjointness condition, the straightforward algorithm appears to require about
$|\myA| |\myB|$ ring operations and tests of disjointness. 

In many cases, we fortunately can do better by applying the principle of
inclusion and exclusion. The basic idea is to compute the sum over pairs $(A,
B)$ with $A \cap B =  \emptyset$ by subtracting the sum over pairs with $A \cap
B =  X \neq \emptyset$ from the sum over pairs with no constraints. For a
precise treatment, it is handy to  denote by $N$ the union of all the members in
the families $\myA$ and $\myB$,  and extend the functions $\alpha$ and $\beta$
to all subsets of $N$  by letting them evaluate to $0$ outside $\myA$ and
$\myB$, respectively. We also use the Iverson bracket notation: $[P] = 1$ if $P$
is true, and $[P] = 0$ otherwise.  Now, by elementary manipulation, 
\be 
\nonumber	
	\ds{\alpha}{\beta}
	\;& = &\;
	\sum_{A}\sum_{B}\,[A \cap B = \emptyset]\, \alpha(A)\, \beta(B)\\
\nonumber	
	\;& = &\;
	\sum_{A}\sum_{B} 
	\sum_{X}\,(-1)^{|X|}\,[X \subseteq A \cap B]\,\alpha(A)\, \beta(B)\\
\nonumber	
	\;& = &\; 
	\sum_{X}\, (-1)^{|X|} \sum_{A}\sum_{B}\,
	[X \subseteq A]\,[X \subseteq B]\,\alpha(A)\, \beta(B)\\ 
	\;& = &\;
	\sum_{X}\, (-1)^{|X|} 
	\Big(\sum_{A \supseteq X} \alpha(A)\Big)
	\Big(\sum_{B \supseteq X} \beta(B) \Big)\;. \label{eq:ds}
\ee
Here we understand that $A$, $B$, and $X$ run through all subsets of
$N$ unless otherwise specified. Note also that the second equality holds
because every {\em nonempty} set has exactly as many subsets of even
size as subsets of odd size.

To analyze the complexity of evaluating the inclusion--exclusion expression 
(\ref{eq:ds}), 
we define the {\em lower set} of a set family $\myF$, denoted by
$\downset\myF$, as the family consisting of all the sets in $\myF$
and their subsets. We first observe that in (\ref{eq:ds}) it suffices to
let $X$ run over the intersection of $\downset\myA$ and $\downset\myB$, for any
other $X$ has no supersets in $\myA$ or in $\myB$. Second, we observe that the
values 
\bes
	\hat{\alpha}(X) \doteq \sum_{A \supseteq X} \alpha(A)\;,
\ees
for all $X \in \downset\myA$, can be computed in 
a total of $|\downset\myA|\, n$ ring and set operations, as follows. 
Let $a_1, a_2,\ldots, a_n$ be the $n$ elements of $N$. For any 
$i = 0, 1, \ldots, n$ and $X \in \downset\myA$ define 
$\hat{\alpha}_i(X)$ as the sum of the $\alpha(A)$ over all sets 
$A \in \downset\myA$
with $A \supseteq X$ and 
$A\cap \{a_1,a_2,\ldots,a_i\} = X \cap \{a_1,a_2,\ldots,a_i\}$. 
In particular, $\hat{\alpha}_n(X) = \alpha(X)$ and 
$\hat{\alpha}_0(X) = \hat{\alpha}(X)$.
Furthermore, by induction on $i$ one can prove the recurrence  
\bes
	\hat{\alpha}_{i-1}(X) 
	= 
	[a_i \not\in X]\, \hat{\alpha}_{i}(X) 
	+ [X\cup\{a_i\} \in \downset\myA]\, \hat{\alpha}_{i}(X\cup\{a_i\})\;; 
\ees
for details, see closely 
related recent work on trimmed zeta transform and Moebius inversion
\cite{BHKK-stoc,BHKK-stacs}. 
Thus, for each $i$, the values $\hat{\alpha}_i(X)$ for all $X \in \downset\myA$ 
can be computed with $|\downset\myA|$ ring and set operations.

We have shown the following. 

\begin{theorem}\label{thm:ds}
The Disjoint Sum problem can be solved with
$\nO\big(n\, (\,|\downset\myA| + |\downset\myB|\,)\big)$ ring and set operations, 
and with a storage for $\nO\big(\,|\downset\myA| + |\downset\myB|\,\big)$
ring elements, 
where $n$ is the number of distinct elements covered by the members of $\myA$
and $\myB$. 
\end{theorem}

\section{Paths}

Consider paths in an undirected graph with vertex set $V$ and edge set $E$.
Define a {\em $k$-edge path} as a sequence of $k+1$ distinct vertices $v_0 v_1
\cdots v_{k}$ such that the adjacent vertices  $v_{i-1}$ and $v_i$ are connected
by an edge $v_{i-1}v_i$ in $E$, for $i=1,2,\ldots,k$. We call the set $\{v_0,
v_1,\ldots, v_k\}$ the {\em support} of the path and $v_0$ and $v_k$ the {\em
ends} of the path. For any vertex $v$ and a subset of $j$ vertices $S\subseteq
V$, let $p_j(S, v)$ denote the number of $j$-edge paths with an end $v$ and
support $S\cup\{v\}$. Clearly, the values  can be computed by dynamic
programming using the recurrence 
\bes
	p_0(S, v) = [S = \emptyset]\;,\quad
	p_j(S, v) = \sum_{u \in S} 
	p_{j-1}(S\setminus\{u\}, u)\,[uv\in E]
	\quad \textrm{for } j>0\;.
\ees
Alternatively, one may use the inclusion--exclusion formula 
\cite{Karp,KohnGottliebKohn}
\bes
	p_j(S, v) = \sum_{Y \subseteq S} (-1)^{|S\setminus Y|}\,  w_j(Y, v)\;,
\ees
where $w_j(Y, v)$ is the number of $j$-edge walks starting from $v$ and visiting
some vertices of $Y$, that is, sequences $u_0 u_1 \cdots u_j$ with $u_0 = v$,
each $u_{i-1}u_i \in E$, and $u_1, u_2, \ldots, u_j \in Y$. Note that for any given 
$Y$, $v$, and $j$, the term $w_j(Y, v)$ can be computed in time 
polynomial in $n$.
Using either of the above two formulas, the values $p_j(S, v)$ for all  
$v \in V$ and sets $S \subseteq V\setminus\{v\}$ of size $j$, can be computed in
time $\Ostar\big(\binsum{n}{j}\big)$; 
here and henceforth, $\binsum{q}{r}$ denotes the sum of the binomial
coefficients  ${q \choose 0} + {q \choose 1} + \cdots + {q \choose r}$. 
In particular, the number of $k$-edge paths in
the graph is obtained as the sum of $p_k(S, v)$ over all 
$v \in V$ and $S \subseteq V\setminus\{v\}$ of size $k$, in time 
$\Ostar\big(\binsum{n}{k}\big)$.

\sloppy
However, meet-in-the-middle yields a much faster algorithm.   Assuming for
simplicity that $k$ is even, the path has a mid-vertex, $v_{k/2}$, at which
the path uniquely decomposes into two $k/2$-edge paths, namely $v_0v_1\cdots
v_{k/2}$ and $v_{k/2}v_{k/2+1}\cdots v_k$, with almost disjoint supports.  Thus,
the number of $k$-edge paths is obtained as the sum of the products
\bes
	p_{k/2}(S, v)\,p_{k/2}(T, v)\Big/2
\ees
over all vertices $v \in V$ and disjoint pairs of subsets $S, T \subseteq
V\setminus\{v\}$ of size $k/2$. 
Applying Theorem~\ref{eq:ds}, once for each $v\in V$, with 
$\myA \doteq \myB \doteq \{S \subseteq V\setminus\{v\} : |S| = k/2\}$ and
$\alpha \doteq \beta \doteq p_{k/2}$ gives the following.

\begin{theorem}\label{thm:path}
The $k$-edge paths in a given graph on  
$n$ vertices can be counted in time $\Ostar\big({n \choose k/2}\big)$.  
\end{theorem}

In the remainder of this section we present a polynomial-space variant of the
above described algorithm. Let the mid-vertex $v$ be fixed. Then 
the task is to compute, for each  $X \subseteq V\setminus\{v\}$ 
of size at most $k/2$, the sum 
\bes
	\sum_{S \supseteq X} p_{k/2}(S, v)
	& = & 
	\sum_{S \supseteq X} 
	\sum_{Y \subseteq S} (-1)^{|S\setminus Y|}\,  w_{k/2}(Y, v)
\ees 
in space polynomial in $n$ and $k$. If done in a straightforward manner, the
running time, ignoring polynomial factors, becomes proportional to the number of triplets $(X, S, Y)$ 
with $X, Y \subseteq S \subseteq V\setminus\{v\}$ and $|S| = k/2$. 
This number is ${n-1 \choose k/2} 2^{k}$ because there are 
${n-1 \choose k/2}$ choices for $S$ and for any fixed $S$, there are 
 $2^{k/2}$ choices for $X$ and $2^{k/2}$ choices for $Y$.

A faster algorithm is obtained by reversing the order of summation: 
\bes
	\sum_{S \supseteq X} p_{k/2}(S, v)
	& = & 
	\sum_{Y} w_{k/2}(Y, v)
	\sum_{S} (-1)^{|S\setminus Y|}\, [X, Y \subseteq S]\\
	& = & 
	\sum_{Y} w_{k/2}(Y, v)\,(-1)^{k/2-|Y|}\,
	{n-|X\cup Y| \choose k/2-|X\cup Y|}\;;
\ees
here $Y$ and $S$ run through all subsets of $V\setminus\{v\}$ of  size at most
$k/2$ and exactly $k/2$, respectively.  The latter equality holds because $S$ is
of size $k/2$ and contains $X\cup Y$. It remains to find in how many ways one
can choose the sets $X$ and $Y$ such that the union $U \doteq
X\cup Y$ is of size at most $k/2$. This number is  
\bes
	\sum_{s=0}^{k/2} {n-1 \choose s} 3^{s} 
 	\,& \leq &\, 
	\frac{3}{2}{n-1 \choose k/2} 3^{k/2}\;,
\ees
because
there are ${n-1 \choose s}$ ways to choose $U$ of size $s$, 
and one can put each element in $U$ either to $X$ or $Y$ or both.

\begin{theorem}\label{thm:polypath}
The $k$-edge paths in a given graph on  
$n$ vertices can be counted in time 
$\Ostar\big(3^{k/2} {n \choose k/2}\big)$
in space polynomial in $n$ and $k$.  
\end{theorem}

\section{Set Packing}

Next, consider packings in a set family $\myF$ consisting of
subsets of a universe $N$. We will assume that each member of $\myF$ is of
size $m$.  A {\em $k$-packing}  in $\myF$ is a set of $k$ mutually disjoint
members of  $\myF$. The members $F_1, F_2, \ldots, F_k$ of a $k$-packing  can be
ordered in $k!$ different ways to an  {\em ordered $k$-packing} $F_1 F_2\cdots
F_k$.   Define the {\em support} of the ordered $k$-packing as the union of its
members.   For any $S \subseteq N$, let $\pi_j(S)$ denote the number of
ordered $j$-packings in $\myF$ with support $S$. The values can be
computing by dynamic programming using the  recurrence
\bes
	\pi_0(S) = [S = \emptyset]\;,\quad
	\pi_j(S) = \sum_{F \subseteq S} 
	\pi_{j-1}(S\setminus F)\, [F\in\myF]
	\quad \textrm{for } j>0\;.
\ees
Alternatively, one may use the inclusion--exclusion formula 
\bes
	\pi_j(S) = \sum_{Y \subseteq S}(-1)^{|S\setminus Y|} 
	\bigg(\sum_{F \subseteq Y} [F \in \myF]\bigg)^j\, 
\ees
(here we use the assumption that every member of $\myF$ is of size $m$)
\cite{BHK,BHKK-stoc}. Using the inclusion--exclusion formula,  the values 
$\pi_{j}(S)$  for all $S\subseteq N$ of size $mj$ can be computed in  time 
$\Ostar\big(\binsum{n}{mj}\big)$, where $n$ is the cardinality of $N$; a
straightforward implementation of the  dynamic programming algorithm yields the
same bound, provided that $m$ is a constant. In particular, the number of
$k$-packings in $\myF$ is obtained as the sum of $\pi_k(S)\big/ k!$ over all
$S\subseteq N$ of size $mk$,  in time $\Ostar\big(\binsum{n}{mk}\big)$.

Again, meet-in-the-middle gives a much faster algorithm. 
Assuming for simplicity that $k$ is
even, we observe that the ordered  $k$-packing decomposes uniquely into two
ordered $k/2$-packings $F_1 F_2 \cdots F_{k/2}$ and $F_{k/2+1}F_{k/2+2}\cdots F_k$ 
with disjoint supports.  
Thus the number of ordered $k$-packings in $\myF$ is
obtained as the sum of the products
\bes
   \pi_{k/2}(S)\,\pi_{k/2}(T)\Big/ 2
\ees
over all disjoint pairs of subsets $S, T \subseteq N$ of size $mk/2$.
Applying Theorem~\ref{thm:ds} with 
$\myA \doteq \myB \doteq \{S \subseteq N : |S| =mk/2\}$ and
$\alpha \doteq \beta \doteq \pi_{k/2}$ gives the following.

\begin{theorem}\label{thm:pack}
The $k$-packings in a given family of $m$-element subsets of an
$n$-element set can be counted in time $\Ostar\big({n \choose mk/2}\big)$.  
\end{theorem}

We next present a polynomial-space variant. The task is, in essence,  
to compute for each  $X \subseteq N$ of size at most $mk/2$ the sum 
\bes
	\sum_{S \supseteq X} \pi_{k/2}(S)
	& = & 
	\sum_{S \supseteq X} 
	\sum_{Y \subseteq S}(-1)^{|S\setminus Y|} 
	\bigg(\sum_{F \subseteq Y} [F \in \myF]\bigg)^{k/2}
\ees 
in space polynomial in $n$, $k$, and $m$.

As with counting paths in the previous section, 
a faster than the straightforward algorithm 
is obtained by reversing the order of summation: 
\bes
	\sum_{S \supseteq X} \pi_{k/2}(S)
	& = &
	\sum_{Y} \bigg(\sum_{F \subseteq Y} [F \in \myF]\bigg)^{k/2}
	\sum_{S} (-1)^{|S\setminus Y|} [X, Y \subseteq S]\\
	& = &
	\sum_{Y} \bigg(\sum_{F \subseteq Y} [F \in \myF]\bigg)^{k/2}\,
	(-1)^{k/2-|Y|}\,
	{n-|X\cup Y| \choose mk/2-|X\cup Y|}\,;
\ees
here $Y$ and $S$ run through all subsets of $N$ of 
size at most $mk/2$ and exactly $mk/2$, respectively. 	
It remains to find the number of triplets $(X, Y, F)$ 
satisfying $|X\cup Y| \leq mk/2$, $|F|=m$, and $F \subseteq Y$. 
This number is 
\be \label{eq:bound}
	\sum_{s=m}^{mk/2} {n \choose s} {s \choose m} 2^m 3^{s-m} 
\nonumber	& < &
	\frac{3}{2}\,{n \choose mk/2}{mk/2 \choose m} 2^m 3^{mk/2-m}\\
	& \leq & 
	\frac{3}{2}\,{n \choose mk/2} 5^{mk/2}\;,
\ee 
because there are ${n \choose mk/2}$ choices for the union 
$U \doteq X\cup Y$ of size $s$, within which there are ${s \choose m}$ 
choices for $F$;
the elements in $F$ can be put to only $Y$ or to both $X$ and $Y$, whereas each
of the
remaining $s - m$ elements in $U$ is put to either $X$ or $Y$ or both.   

\begin{theorem}\label{thm:polypack}
The $k$-packings in a given family of $m$-element subsets of an
$n$-element set can be counted in time 
$\Ostar\big(5^{mk/2} {n \choose mk/2}\big)$ in
space polynomial in $n$, $k$, and $m$.  
\end{theorem}

We remark that the upper bound (\ref{eq:bound}) is rather crude for small values
of $m$. In particular, provided that $m$ is a constant, we can replace the
constant $5$ by $3$. 

\begin{corollary}\label{cor:m2}
The $k$-packings in a given family of\/ $2$-element subsets of an
$n$-element set can be counted in time 
$\Ostar\big(3^{k} {n \choose k}\big)$ in
space polynomial in $n$ and $k$.  
\end{corollary}

\end{document}